# Observation of Landau level–like quantizations at 77 K along a strained-induced graphene ridge

Lin He[1*], Hui Yan[1], Yi Sun[1], Jia-Cai Nie[1], and Moses H. W. Chan[2]

[1] Department of Physics, Beijing Normal University, Beijing, 100875, People's Republic of China

[2] The Center for Nanoscale Science and Department of Physics, The Pennsylvania State University, University Park, Pennsylvania 16802-6300, USA

Recent studies show that the electronic structures of graphene can be modified by strain and it was predicted that strain in graphene can induce peaks in the local density of states (LDOS) mimicking Landau levels (LLs) generated in the presence of a large magnetic field. Here we report scanning tunnelling spectroscopy (STS) observation of nine strain-induced peaks in LDOS at 77 K along a graphene ridge created when the graphene layer was cleaved from a sample of highly oriented pyrolytic graphite (HOPG). The energies of these peaks follow the progression of LLs of massless 'Dirac fermions' (DFs) in a magnetic field of 230 T. The results presented here suggest a possible route to realize zero-field quantum Hall-like effects at 77 K.

The physics of 'Dirac fermions' (DFs) in condensed-matter has received a great deal of attention, in part because it has opened up a new area of fundamental science and also for its long-term potential applications [1-7]. In monolayer graphene, the DFs move as if they were massless at the Fermi velocity $v_F \sim 10^6$ m/s. The quantum-relativistic nature of the massless DFs was obtained through the appearance of the unusual Landau levels (LLs) energy spectrum, which consists of a sequence of levels with square-root dependence on both magnetic field and level index $n$ and a unique level $n = 0$ at an energy that is pinned to the Dirac point $E_D$ [5]. Upon cleaving a highly oriented pyrolytic graphite (HOPG) sample, strains can be induced in the resultant graphene layer forming folded-over structure, wrinkles, and ridges [8,9]. These surface structures provide model systems for the study of the coupling between the graphene and the underlayers [10-14]. Recently, it was shown that strain induced gauge field can be used to alter and tailor the electronic structure of graphene. This opens a possible avenue towards all-graphene electronics [15-18]. Defects in the form of nanobubbles are also found on chemical vapor deposited graphene layer [17]. Levy and co-authors observed three peaks in scanning tunnelling spectroscopy (STS) spectra at positive voltage bias measured at 7.5 K under zero magnetic field and show evidence that these peaks mimic LLs of massless DFs generated in the presence of large magnetic field [17]. Here, we present STS results at 77 K showing 9 distinct peaks in the tunnelling spectra along a graphene ridge on HOPG produced during the cleaving process. The energy values of these 9 peaks follows the sequence of LLs of massless DFs in a magnetic field. The deviations from the expected values of energies of the peaks are also observed. Such deviations are reasonable since these peaks are not results of Landau level quantizations in a real magentic field but the consequence of the distortion of the local density of states (LDOS) induced by strain.

The scanning probe microscope (STM) system used in the experiments is an ultrahigh vacuum four-probe STM from UNISOKU. All the STM and STS measurements were carried out at liquid-nitrogen temperature in a constant-current scanning mode. The samples were cleaved in air, and the scaning tunnelling microscope tips were obtained from mechanically cut Pt (80%)/ Ir (20%) wire. The tip-sample distance was set by a tunnelling current of 49 pA with sample bias voltage of 1.39 V unless otherwise specified. The d$I$/d$V$ measurements were carried out with a standard lock-in technique using a 957 Hz a.c. modulation of the bias voltage.

Figure 1 shows a scanning tunnelling microscopy (STM) topograph of a freshly cleaved graphite surface with a graphene ridge from A to B. The height of the ridge is about 10 nm and the length from A to B is about 60 nm. Due to the curvature, it is difficult to obtain atomic-resolution images along the ridge of the graphene layer. Atomic resolution image, which is composed of hexagonal lattices with a periodicity of 0.246 nm, was obtained about 500 nm from the ridge. This is shown in the inset of the

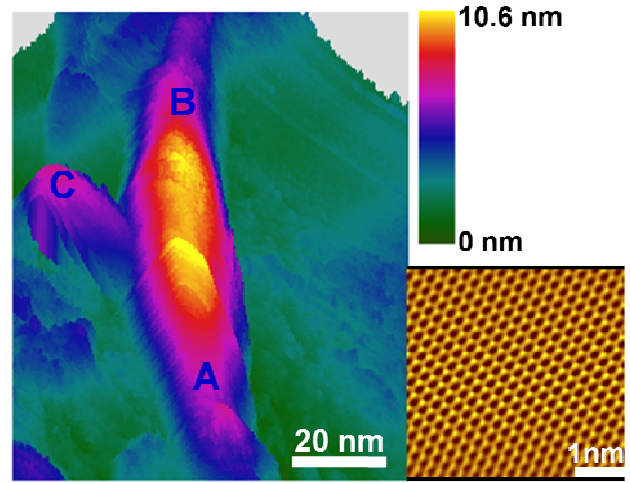

FIG. 1. STM image of a graphene ridge on HOPG observed at 77 K. The length of the ridge is about 60 nm (between A and B) and the width is about 15 nm. The inset in the lower panel shows the STM image on a flat graphite surface about 500 nm from the graphene ridge. STS measurements along the ridge show a series of 9 peaks in d$I$/d$V$ curve corresponding to Landau levels at 230 T. No d$I$/d$V$ peak is seen at the elevated region labeled as C. It may be a dirt particle, or a broken graphite remnant.

figure. While atomic resolution image was not possible, STS spectra of the local electronic structure along the ridge were possible. A d$I$/d$V$ vs bias voltage ($V_{bias}$) spectrum measured at 77 K at one point on the ridge between A and B and of the flat region far from the ridge is shown in Figure 2. The tunnelling spectrum gives direct access to the LDOS of the surface at the position of the STM tip. While the spectrum in the flat region (inset) show no discernible structure, nine peaks are clearly resolved in the spectrum measured on the ridge. Spectra measured in the other elevated region shown in Figure 1 and labeled as C are similar to that found in the flat region, *i.e.*, with no evidence of any peak. This indicates that the elevated region C may be a dirt particle, or a broken graphitic remnant. The STS or d$I$/d$V$ data taken on the graphene ridge can be fitted with a sequence of Gaussian peaks (shown in green) plus a simple polynomial background. These peaks are labelled with integer indices.

In order to ascertain the reproducibility of the results, a total of more than 100 d$I$/d$V$ spectra were measured at different positions along the ridge between A and B with two different STM tips over a time interval of 12 hours at night. For clarity, only 13 spectra measured at different locations along the ridge are shown in Figure 3(a). The spectrum shown in Figure 2 is reproduced in Fig. 3(a) as the third spectrum from the bottom. All the experimental spectra show similar peak structure with small variations in peak spacings and amplitudes. The peak positions are not strictly periodic. This eliminates quantum confinement as a possible origin of the peaks [15]. Following reference 17, we attribute these peaks to the strain-induced LDOS along the graphene ridge and compare the progression of these peaks with LLs splitting of graphene under a magnetic field [16]. This means the peak positions $E_n$ (= e$V_{bias}$, where $V_{bias}$ is the bias voltage) should follow

$$E_n = E_D + \text{sgn}(n) v_F \sqrt{2eB\hbar|n|}, \quad n = 0, \pm1, \pm2, ..., \quad (1)$$

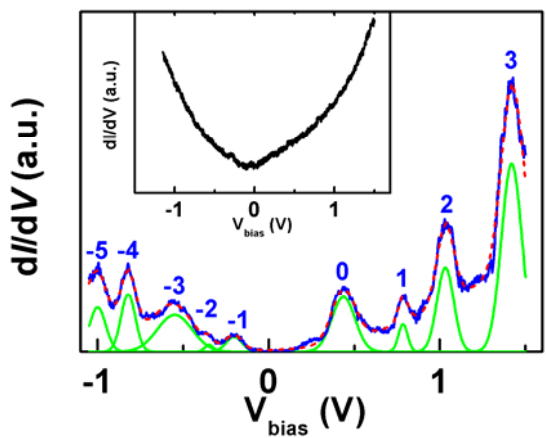

FIG. 2. Tunnelling spectrum on graphene ridge. A typical d$I$/d$V$-$V$ curve (blue solid curve) of the graphene ridge measured at 77 K. The red dashed curve is the sum of a sequence of Landau level (Guassian) peaks (shown in green) and a simple polynomial background. The index $n$ is shown above the peaks of the spectrum. The inset shows the d$I$/d$V$-$V$ curve obtained on a flat graphite surface about 500 nm from the graphene ridge.

where $\hbar$ is the Planck's constant, $B$ the magnetic field, and $E_D$ the position in energy of the Dirac point [5]. In this expression, both positive and negative $n$ are expected to appear symmetrically about the Dirac point, corresponding to the $n = 0$ state.

In Figure 3(b), we plot $E_n$ in units of e$V$ as a function of sgn$(n)|n|^{1/2}$ for all the peaks that appear in the 13 spectra shown in Figure 3(a). A reasonable linear depedence is found yielding $E_D \sim 330$ meV and a pseudo-magnetic field $B$ of 230 T. The fitted value of $E_D$ at 330 meV above the Fermi energy is consistent with the Dirac point of strained graphene observed in reference 17. The strain-induced pseudo-magnetic flux of a corrugated graphene sheet was calculated [19]. The flux per corrugation was found to be $\Phi = (\beta h^2/la)\Phi_0$. Here $h$ is the height, $l$ the length of corrugation, $a$ is on the order of the C-C bond length, and $\Phi_0$ the flux quantum [19]. The parameter $\beta$, which represents the change in the hopping amplitude between nearest neighbor carbon atoms to bond length, has a typical magnitude of $2 < \beta < 3$ for graphene [17,19]. When we make use of this relation for our graphene ridge with $h \sim 10$ nm, $l \sim 60$ nm, and the width of the ridge to be 15 nm, the strain induced pseudo-magnetic field is found to be only 40 T. The experimental result of 230 T indicates that the strain in our graphene ridge is larger than that of a well ordered corrugated graphene structure of similar size. Our results as shown in Figure 3(b) nevertheless demonstrate that the strain in graphene does mimic the effect of a real perpendicular magnetic field on the LDOS.

While equation 1 provides a reasonable description of the 9 peaks of the 13 spectra, Figure 3(b) also showed that the peak positions of the different spectra exhibit obvious deviations from linear relation as predicted for real Landau levels. These deviations are highlighted in a deviation plot shown in Figure 3(c). Figures 3(b) and 3(c) show two different kinds of deviations. Firstly, the values of $E_n$ at a particular $n$ of the (13) different spectra obtained at different points along the ridge are dissimilar from each other. These deviations reflect, in addition to instrumental resolution, also the variation in the strain at different positions along the ridge. In addition, the positions or energies of the peaks appear to show systematic deviations from the prediction of equation 1. Specfically, in the spectrum shown in Figure 2, the spacing between peaks labelled by indices 0 and -1 (~ 0.64 V) is nearly twice of that between 0 and 1 (~ 0.35 V). This asymmetry in the relative spacings of these three peaks is also found in all the spectra shown in Figure 3(a) and reflected in Figure 3(c) in the deviation plot. Figure 3(c) shows that peaks of all the (13) spectra labelled by index 0 lie above zero and the peaks labelled by index 1 lie below. Interestingly, the ratio of the spacing between peaks labelled by indices 0 and 1 and that between 0 and -1 is about the square-root of 1/3 (~ 0.58) rather than 1 as predicted by equation (1). Very recently, the states with Landau level fractional filling factors were observed in top layer of multilayer epitaxial graphene by STM and STS studies at temperature as low as 10 mK [14]. However, we note that the deviation that appear to favor the '1/3' fraction cannot be related to fractional quantum Hall effects observed in pristine and un-strained graphene [20-24]. A more thorough theoretical analysis of the LL-like sequence with the systematic deviation as shown in Fig. 3 will likely reveal interesting new physics.

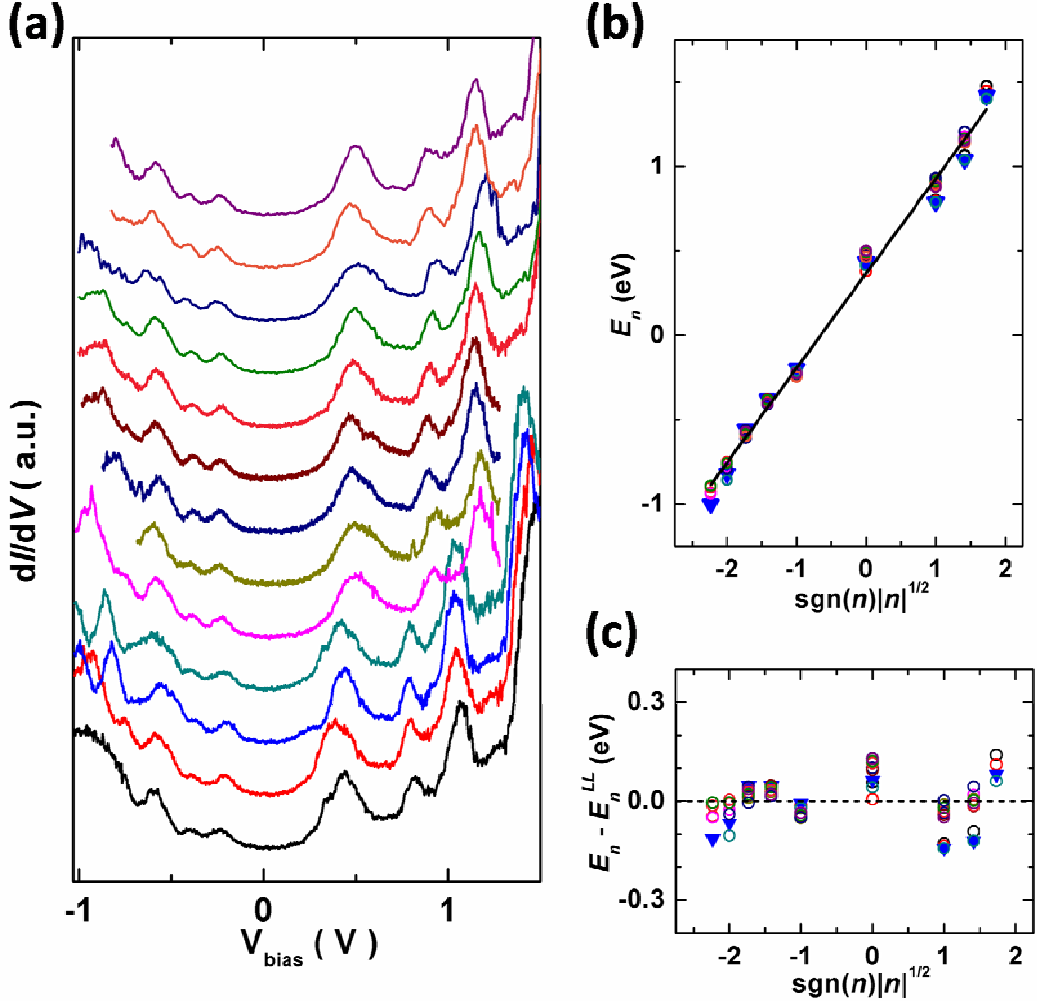

FIG. 3. Tunnelling spectra and Landau level-like quantizations along the graphene ridge. (a), 13 $dI/dV$-$V$ curves of the graphene ridge measured from position A to B. The spectrum shown in Figure 2 is reproduced as the third spectrum from the bottom. (b), The energy of Landau level-like peaks deduced from the spectra from panel (a) as a function of $sgn(n)|n|^{1/2}$. Different colors correspond to data measured at different positions along the ridge. The solid blue triangles label the positions or energies of the peaks shown in the spectrum of Fig. 2. The solid lines are the linear fit of the experimental data to Eq. (1). The fitted pseudo-magnetic field is 230 ± 20 T. The deviations of the peak positions, $E_n$, from the expected LLs, $E_n^{LL}$, are highlighted in a deviation plot shown in (c).

There are obvious similarities between our results and those presented in reference 17. The peaks in their tunnelling spectra arose from strained graphene nanobubbles with width of about 4 to 10 nm and height of 0.3 to 2.0 nm. In their tunnelling spectra, three peaks at positive bias were clearly observable and attributed to 0, 1, and 2 LLs, the peaks at negative bias are not as clear (as shown in the supporting materials of reference 17). In our experiment, the observation of 9 distinct peaks removes any doubt for the interpretation of strain-induced Landau level-like quantizations in our graphene ridge. Additionally, the deviations from the expected values of energies of the peaks are also observed. Such deviations are expected since these peaks are not results of Landau level quantizations in a real magentic field but the consequence of the distortion of the LDOS induced by strain.

To conclude, we observed Landau level-like quantizations in a strained graphene sheet at 77 K. The energies of these peaks follow globally the progression of LLs of massless DFs in a magnetic field of 230 T. This result suggests the possibility of the observation of zero-field quantum Hall-like effect by transport measurements in a strained graphene system [15,16,19,25] at 77 K. Furthermore, the deviations from the expected values of LLs' energies of the peaks are also observed, which indicates that more detailed theoretical studies are needed to fully understand the strain induced Landau level-like quantizations in graphene.

We are grateful to Jun Zhu for helpful discussions. This work was supported by the National Natural Science

Foundation of China (Grant Nos. 10974019 and 11004010), the Fundamental Research Funds for the Central Universities, and the Penn State MRSEC under NSF grant DMR-0820404.

*Email: helin@bnu.edu.cn.

55